\documentclass[aps,prl,floats, twocolumn,showpacs,superscriptaddress]{revtex4}

\usepackage{graphicx,epsfig}
\usepackage{times}
\usepackage{graphics,dcolumn,bm,epic, eepic,fleqn,float}
\usepackage{amssymb,amsmath,multirow,rotate,color}

\begin{document}

\title{Stability of Boolean Multilevel Networks}

\author{Emanuele Cozzo}

\affiliation{Institute for Biocomputation and Physics of Complex
Systems (BIFI), University of Zaragoza, Zaragoza 50018, Spain}

\author{Alex Arenas}

\affiliation{Institute for Biocomputation and Physics of Complex
Systems (BIFI), University of Zaragoza, Zaragoza 50018, Spain}

\affiliation{Departament d'Enginyeria Inform\`atica i Matem\`atiques, Universitat Rovira i Virgili, 43007 Tarragona, Spain}

\author{Yamir Moreno}

\affiliation{Institute for Biocomputation and Physics of Complex
Systems (BIFI), University of Zaragoza, Zaragoza 50018, Spain}

\affiliation{Department of Theoretical Physics, University of 
Zaragoza, Zaragoza 50009, Spain}

\date{\today}

\begin{abstract}
The study of the interplay between the structure and dynamics of complex multilevel systems is a pressing challenge nowadays. In this paper, we use a semi-annealed approximation to study the stability properties of Random Boolean Networks in multiplex (multi-layered) graphs. Our main finding is that the multilevel structure provides a mechanism for the stabilization of the dynamics of the whole system even when individual layers work on the chaotic regime, therefore identifying new ways of feedback between the structure and the dynamics of these systems. Our results point out the need for a conceptual transition from the physics of single layered networks to the physics of multiplex networks. Finally, the fact that the coupling modifies the phase diagram and the critical conditions of the isolated layers suggests that interdependency can be used as a control mechanism.
\end{abstract}


\maketitle

Nearly four decades ago, Random Boolean Networks (RBNs) were introduced as a way to describe the dynamics of biochemical networks \cite{kauffman71,Bornholdt05,aldana,Bornholdt00,Klemm05a,Klemm05b,davidich}. RBNs \cite{aldana,drossel08} consider that each gene of a genetic regulatory network is a node of a directed graph, the direction corresponding to the effect of one gene on the expression of another. The nodes can be in one of two states: they are either \textit{on} (1) or \textit{off} (0) - i.e. in the case of a gene its target protein is expressed or not. The system so composed evolves at discrete time steps. At each time step nodes are updated according to a boolean rule assigned to each node that is a function of its inputs. Notwithstanding the high simplicity of RBNs models, they can capture the behavior of some real regulatory networks \cite{Li} allowing for the study of several dynamical features, above all their critical properties. However, although some coupled Boolean networks have been investigated \cite{coupled1,coupled2}, the vast majority of works has considered RBNs as {\em simplex} networks, in which a single graph is enough to represent all the interactions a given gene is involved in. 

\begin{figure}[!t]
\centering
\includegraphics[width=2.5in,angle=0]{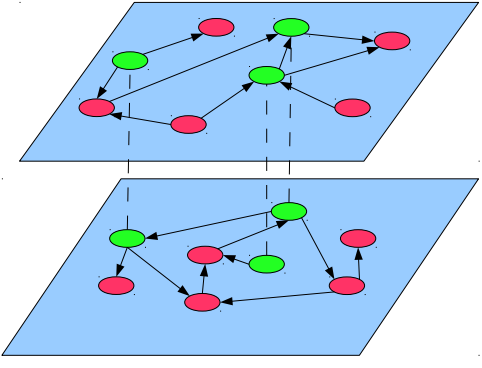}
\caption{(color online) The multiplex network is built up by randomly connecting $N$ nodes per layer. With probability $\sigma$, each of the $N$ nodes can be present in both layers. Therefore, the total number of {\em different nodes} in the system is $\tilde{N}=(2-\sigma)N$. In the example of the figure, the whole system is made up of $\tilde{N}=13$ nodes, of which $3$ are present in the two layers and there are $5$ additional nodes per layer, therefore $N=8$ and $\sigma=3/8$.}
\label{fig1}
\end{figure}

The previous description implicitly assumes that all biochemical signals are equivalent and then collapses information from different pathways. Actually,  in cellular biochemical networks, many different signaling channels do work in parallel \cite{bioessays}, i.e., the same gene or biochemical specie can be involved in a regulatory interaction, in a metabolic reaction or in another signaling pathway. Here, we introduce a more accurate set up for the topology of biochemical networks by considering that different operational levels (pathways) are interconnected layers of interaction. In terms of graphs, this topology is more consistent with a multiplex network \cite{mucha10,koreans12} (see Fig.\ \ref{fig1}) in which each level would represent the different signaling pathways or channels the element participates in. On the other hand, accounting for the multilevel nature of the system dynamics also represents a point of interest by itself, as this allows to inspect what are the consequences of new ways of interdependency between the structure and the dynamics. In this sense, the dynamics we inspect is general enough so as to serve as a null model for many other complex dynamical processes. 

In this paper, we study the stability of Boolean networks defined at multiple topological layers. In particular, we inspect a Boolean multiplex network model, in which each node participates in one or more layers of interactions, being its state in a layer constrained by its own state in another layer. Therefore, we focus on the case of canalizing rules. Boolean functions are canalizing if whenever the canalizing variable takes a given value, the canalizing one, the function always yields the same output. Capitalizing on a semi-annealed approximation, we analytically and numerically study the conditions defining the stability of the aforementioned system. By doing so, we show that the interdependency between the layers can be enough to either stabilize the different levels or the whole system. Remarkably, this also happens for parameter values where the sub-systems, if isolated, were unstable.

Let us first define in mathematical terms the structure of the multiplex network of $N$ nodes per layer and $M$ layers in Fig.\ \ref{fig1}, which can be fully encoded in two objects \cite{note1}. First, we have the $N$x$M$ incidence matrix $B_{i\alpha}$, whose elements are 1 if node $i$ appears in layer $\alpha$ and 0 otherwise. 
Secondly, we introduce an adjacency tensor, $A_{ij\alpha}$, whose elements are 1 if there is a link between nodes $i$ and $j$ in layer $\alpha$ and 0 otherwise. With these two basic objects, one can generalize the different descriptors used in simplex networks. For instance, the {\em total degree} of node $i$ will be $K_i=\sum_{j\alpha}A_{ij\alpha}=\sum_{\alpha}K_{i\alpha}$, where $K_{i\alpha}$ is the degree of node $i$ in layer $\alpha$. Moreover, the multilevel structure gives rise to new topological metrics that are not defined in single-layered networks. We define the \textit{multiplexity degree} of a node as the number of layers in which it appears as $\kappa_i=\sum_\alpha B_{i\alpha}$. Note that the number of different nodes in the multiplex will be then $\tilde{N}=NM- \sum_i(\kappa_i -1)$.

Next, let us consider a state vector 
\begin{equation}
\mathbf{\tilde{x}}(t)=(\tilde{x}_1(t),...,\tilde{x}_{\tilde{N}(t)}),
\label{statevector}
\end{equation}
where $\tilde{x}_i(t)\in\{0,1\}$ and a set of update functions such that
\begin{equation}
\tilde{x}_i(t)=\tilde{f}_i({\tilde{x}}_{j \in \Gamma_{\alpha}^{\mbox{\tiny{in}}}(i)}(t-1)).
\label{systupfunc}
\end{equation}
\noindent where $\Gamma_{\alpha}^{\mbox{\tiny{in}}}$($i$) refers to all the incoming neighbors $j$ of node $i$ at each layer $\alpha$, with $\alpha = 1 \dots M$. 

Equations\ (\ref{statevector}-\ref{systupfunc}) define a Boolean multilevel (or multiplex) graph. In addition, due to the multiplex nature of the network, we also define a set of update functions for each layer as 
\begin{equation}
x^l_i(t)=f^l_i(({\tilde{x}}_{j \in \Gamma_{l}^{\mbox{\tiny{in}}}(i)}(t-1))).
\label{layerupfunc}
\end{equation}
\noindent where now the arguments of the function are restricted to the specific layer $\alpha=l$.  
Equation\ (\ref{layerupfunc}) governs how each node is updated in each layer. So, Eq.\ (\ref{systupfunc}) can be rewritten as  
%
\begin{equation}
\tilde{x}_i(t)=\tilde{f}_i(f^{1}_i, \dots f^{M}_i),
\end{equation}
%
where $\tilde{f}_i$ is a canalizing function of its inputs. These definitions allow investigating how the stability of the Boolean model is affected by the multilevel structure of the system and by the existence of nodes with different multiplexity degrees. 

We first inspect the dependency of the average sensitivity $s^f$, which has been shown to be a useful order parameter in RBNs \cite{senskaufprl,solelyapunov}, on the multiplexity degree $\kappa_i$. Following \cite{senskaufprl}, we write the activity $a_{j}^{f}$ of the variable $x_j$ in a function $f$ of $K$ inputs as $a_j^f=\frac{1}{2^K}\sum_{\mathbf{x}\in\{0,1\}}\frac{\partial f(\mathbf{x})}{\partial x_j}$,
where $\frac{\partial f(\mathbf{x})}{\partial x_j}=f(\mathbf{x}_{(j,0)})\oplus f(\mathbf{x}_{(j,1)})$ and $\mathbf{x}_{(j,R)}$ represents a random vector $\mathbf{x}\in{0,1}$ with the $j$th input fixed to $R$ and $\oplus$ is the arithmetic addition modulo 2. 
Similarly, assuming that the inputs are also uniformly distributed, the average sensitivity is equal to the sum of the activities, i.e., 
%
$s^f=\sum_{i=1}^K E[\chi[{f(\mathbf{x}\oplus e_i)\neq f(\mathbf{x})}]]=\sum_{i=1}^Ka_j^f$, 
%
where $e_i$ is a zeroes vector with 1 in the $i$-th position, and $\chi[A]$ is an indicator function that is equal to 1 if and only if A is true. 
%
%

To illustrate how multiplexity affects the sensitivity of a node, without loss of generality, we study analytically and numerically  a multiplex network of two layers. Let us denote by $p$ the bias of the Boolean functions, and $\alpha$ and $\beta$ the two respective layers. Due to the multilevel nature of the interaction network, a node $i$ in our model depends on the state of its neighbors in layer $\alpha$ and also on the state of its neighbors in $\beta$ via the auxiliary function $\tilde{f}^\beta$. Suppose that the canalizing state in $\alpha$ and $\beta$ is 1 (the discussion for 0 would be identical). Then, the updating function of $i$ can be written as
%
$\tilde{f}_i(f^\alpha_i,f^\beta_i)=f^\alpha \vee f^\beta$, 
%
being $\vee$ the Boolean operator OR. From the definition of the activities and the previous relation, it follows that
%
$E[a_j^{\tilde{f}}]=2^{-(\kappa_i-1)}2p(1-p)$,
%
 which is different from the value one would obtain in the case of a simple canalizing function. Similarly, for the sensitivity one gets
\begin{equation}
E[\tilde{s}^{\tilde{f}}]=2^{-(\kappa_i-1)}\sum_\alpha E[s^{f^{\alpha}}],
\label{avesensi}
\end{equation}
where $E[s^{f^{\alpha}}]=2p(1-p)K_{\alpha}$ is the expected average sensitivity of a function in layer $\alpha$ if it were isolated. 

Next, we study the stability of the Boolean multiplex system using a semi-annealed approximation \cite{pomerance}. This approach considers the network as a static topological object while the update functions $f^l_i$ ($l=\alpha,\beta$) are assigned randomly at each time step. Thus, we can write the update function for the components of the difference vector $\tilde{\mathbf{y}}(t)=\langle\mid\mathbf{\tilde{x}}(t)-\hat{\tilde{\mathbf{x}}}(t)\mid\rangle$, where $\hat{\tilde{x}}$ is a perturbed replica of $\tilde{\mathbf{x}}$ in which a (small) fraction of the nodes were flipped, yielding
\begin{equation}
\tilde{y}_i(t)=\tilde{q}_i[1-\prod_{j\in \mathit{\Gamma}_i}(1-\tilde{y}_j(t-1))]
\label{diff}
\end{equation}
which is equivalent to the expression derived in \cite{pomerance}, but also taking into account Eq.\ (\ref{avesensi}), with $q_i=2p(1-p)$ for a simplex graph and $\mathit{\Gamma}_i$ being the set of all neighbors of $i$ in all layers. Considering a small perturbation, linearization of Eq.\ (\ref{diff}) around the fixed point solution $\tilde{\mathbf{y}}(t)=\mathbf{0}$ leads to
\begin{equation}
\tilde{y}_i(t+1)\approx 2^{-(\kappa_i-1)}q_i\sum_{\alpha=1}^{M}\sum_{j=1}^N A_{ij\alpha}\tilde{y}_j(t)
\end{equation}
that can be written in matrix form as $\tilde{\mathbf{y}}(t+1)=\sum_{\alpha} Q_\alpha \tilde{\mathbf{y}}(t)$, with $Q_{ij\alpha}=2^{-(\kappa_i-1)}q_iA_{ij\alpha}$. The largest eigenvalue, $\lambda_Q$, of the matrix $Q=\sum_{\alpha}Q_{\alpha}$ governs the stability of the system \cite{pomerance}. It is worth noticing that the latter refers to the stability condition {\em for the whole system} and, given a fixed topology for each layer, it depends on the multiplexity degree \cite{note3}.
\begin{figure}[!t]
\centering
\includegraphics[width=2.5in,angle=-90]{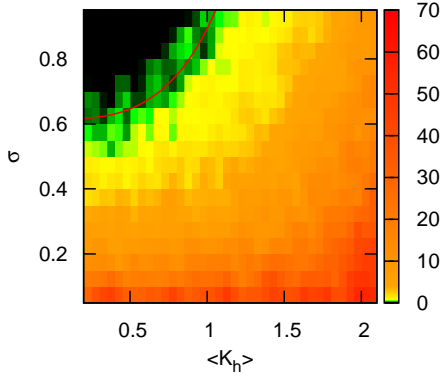}
\caption{(color online) Color-coded average Hamming distance for the whole system with fixed observed connectivity $\langle K_{\mbox{\scriptsize{o}}}\rangle=2.9$ for different values of the hidden connectivity $\langle K_h\rangle$, and the probability for a node in a layer to be present also in the other layer $\sigma$. The network is composed of $N=10^3$ nodes per layer as explained in Fig\ \ref{fig1}. The continuos line is the solution (zeros) of Eq.\ (\ref{criticaltotal}). Simulations were performed for an initial Hamming distance of $0.01$ and the results are averages over $50$ realizations of the network and $300$ random initial conditions.}
\label{fig2}
\end{figure}
For the case of nonuniform $\kappa_i$ we obtain an analogous mean-field approximation to $\lambda_Q$ in \cite{pomerance}, 
\begin{equation}
\lambda_Q\approx \frac{\langle 2^{-(\kappa_i-1)}q_i K_i^{\mbox{\scriptsize{in}}}K_i^{\mbox{\scriptsize{out}}}\rangle}{\langle K\rangle},
\label{lqc}
\end{equation}
where $\langle K\rangle$ is the average degree of the multiplex. Note that the stability of the multiplex depends on $\kappa_i$ and $K_i^{\mbox{\scriptsize{in}}}K_i^{\mbox{\scriptsize{out}}}$, which, in general, are not independent variables $-$ thus, $\tilde{q}_i$ and $K_i$ are anticorrelated. To find the critical condition let $\tilde{P}(\kappa=n)$ be the probability that a node in the whole system has multiplexity degree $n$. This magnitude depends on the same quantity but at the single layer level as $
\tilde{P}(\kappa=n)=\frac{N}{\tilde{N}}\frac{M}{n}P(\kappa=n)$, 
where $P(\kappa=n)$ is the probability that a randomly chosen node of a layer has multiplexity degree $n$. For the average degree of the multiplex we have:
\begin{equation}
\langle K\rangle=\sum_n\frac{\binom{M-1}{n-1}}{\binom{M}{n}}\tilde{P}(\kappa=n)\sum_l\langle K_l\rangle=\frac{N}{\tilde{N}}\sum_l\langle K_l\rangle,
\end{equation}
where $\langle K_l\rangle$ is the average degree of layer $l$. 

Inserting the previous expression into Eq.\ (\ref{lqc}) and considering the case in which there are no correlations between $K^{in}$ and $K^{out}$, one gets,
\begin{equation}
\small{\langle\tilde{q}\rangle\sum_l \langle K_l\rangle-\frac{2(M\langle\tilde{q}\rangle-\langle \kappa\tilde{q}\rangle)}{M-1}\frac{\displaystyle\sum_{l_1<l_2}\langle K_{l_1}\rangle \langle K_{l_2}\rangle}{\sum_l \langle K_l\rangle}=1,}
\label{general}
\end{equation}
with $l_1=1\dots,M$, $l_2=1\dots,M$ and $\langle\tilde{q}\rangle=\sum_{n=1}^{M}\tilde{q}(\kappa=n)P(\kappa=n)$ is the average sensitivity on a layer. It is worth noticing that the first term on the l.h.s. of Eq.\ (\ref{general}) is the expression one would obtain using an annealed approximation. The second term is always positive. Therefore, it captures the stabilizing effects of multiplexity, rightly predicting ordered behavior in regions in which the annealed approximation would not.

\begin{figure}[!t]
\centering
\includegraphics[width=3.2in,angle=0]{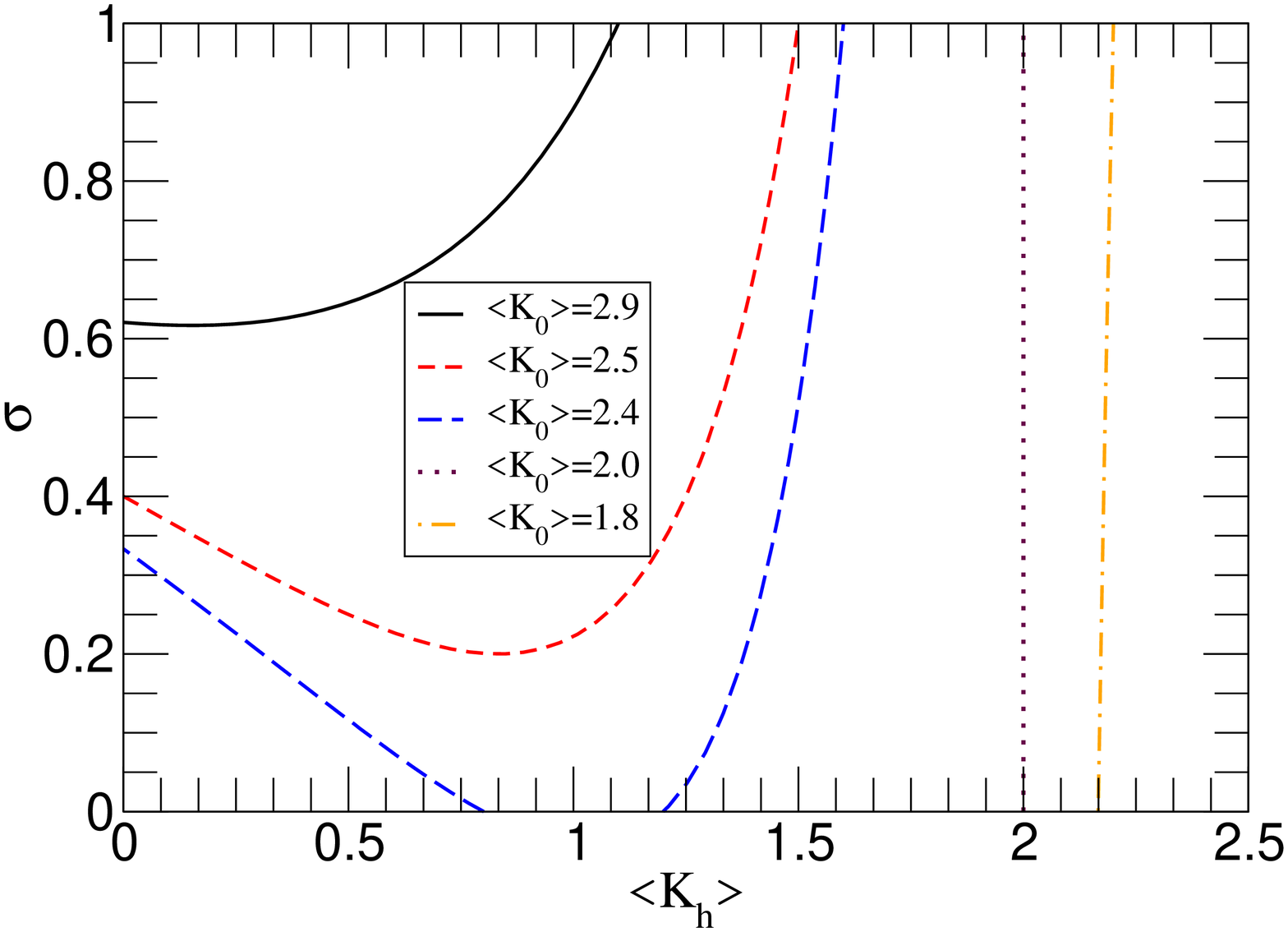}
\caption{(color online) The lines are the solution (zeros) of Eq.\ (\ref{criticaltotal}) for different values of the hidden connectivity $\langle K_{\mbox{\scriptsize{h}}}\rangle$, the observed connectivity $\langle K_{\mbox{\scriptsize{o}}}\rangle$ and the probability of a node belongs to both layers $\sigma$. We have set $q_i=q=\frac{1}{2}$.}
\label{fig3}
\end{figure}

Once we have derived the critical condition for a system made up of an arbitrary number of layers, let us compare the analytical results with numerical simulations for a two-layers system with $q_i=q$. Let $\sigma$ be the probability for a node in a layer to be present also in the other layer, then we have $P(\kappa=2)=\sigma$ and $P(\kappa=1)=1-\sigma$. Besides, for the sake of simplicity, consider that the average connectivity of one layer is observed, $\langle K_{\mbox{\scriptsize{o}}}\rangle$, and fixed (for instance, because one measures it), and that the average connectivity of the other layer is unknown or hidden $\langle K_{\mbox{\scriptsize{h}}}\rangle$. Recalling that the size of the multiplex system is $\tilde{N}=(2-\sigma)N$ $-$where $N$ is the number of nodes per layer$-$, the mean connectivity $\langle K\rangle$ can be written as
%
$\langle K\rangle=\frac{\langle K_{\mbox{\scriptsize{h}}}\rangle+\langle K_{\mbox{\scriptsize{o}}}\rangle}{(2-\sigma)}$, 
%
which leads to the following expression for the critical condition of the two-layers system
\begin{equation}
\frac{2-\sigma}{4}(\langle K_{\mbox{\scriptsize{h}}}\rangle+\langle K_{\mbox{\scriptsize{o}}}\rangle)-(1-\sigma)\frac{\langle K_{\mbox{\scriptsize{h}}}\rangle\langle K_{\mbox{\scriptsize{o}}}\rangle}{\langle K_{\mbox{\scriptsize{h}}}\rangle+\langle K_{\mbox{\scriptsize{o}}}\rangle}=\frac{1}{2q}
\label{criticaltotal}
\end{equation}
that as a function of $\sigma$ and $\langle K_{\mbox{\scriptsize{h}}}\rangle$ gives an hyperbolic critical curve. 

To verify that our analytical calculations are valid, we have performed extensive numerical simulations of the Boolean dynamics on a random multiplex network made up of two layers in which $N$ nodes are randomly connected among them and only a fraction $\sigma$ of them are present on both layers. As it is customarily done, we test the stability of the system by measuring the long-time Hamming distance for different trajectories generated from two close initial states. Figure\ \ref{fig2} shows the results obtained when the mean connectivity $\langle K_{\mbox{\scriptsize{o}}}\rangle$ of a layer is fixed and both $\sigma$ and the mean connectivity of the other layer $\langle K_{\mbox{\scriptsize{h}}}\rangle$ change (the Hamming distance is color coded as indicated). First, we note that the transition from stability to an unstable regime nicely agrees with the theoretical prediction. Secondly, it is worth highlighting a new effect linked to the multi-level nature of the system: the region of low $\langle K_{\mbox{\scriptsize{h}}}\rangle$ and low $\sigma$ is unstable despite the fact that those values of $\langle K_{\mbox{\scriptsize{h}}}\rangle$ would make the hidden layer, in a {\em simplex graph} description, stable. However, due to the low coupling ($\sigma$), the instability of the multiplex is determined by that of the observed layer, the leading one. Admittedly, when increasing the coupling $\sigma$ the stable (hidden) layer is able to stabilize the whole system.

We have further explored the dependency between the stability of the multiplex and the average degrees of both layers. Figure\ \ref{fig3} shows the analytical solution of Eq.\ (\ref{criticaltotal}) for different values of $\langle K_{\mbox{\scriptsize{h}}}\rangle$ and $\langle K_{\mbox{\scriptsize{o}}}\rangle$. The results show a very rich phase diagram. Depending on the values of both connectivities, a double transition from a chaotic regime to an ordered one and again to another chaotic regime is predicted. More interestingly, the transition from the ordered to the disordered regime does not depend on $\sigma$ only when both layers operate at their respective critical points, namely, when $\langle K_{\mbox{\scriptsize{h}}}\rangle=\langle K_{\mbox{\scriptsize{o}}}\rangle=1/q=2$.

\begin{figure}[!t]
\centering
\includegraphics[width=3.2in,angle=0]{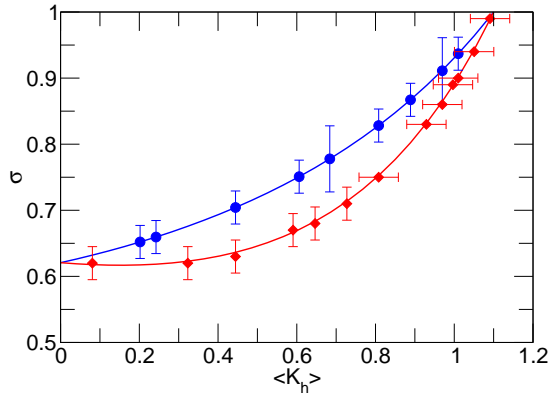} 
\caption{(color online) Critical curves for a network made up of $10^4$ nodes per layer as a function of the probability of a node to be part of both layers $\sigma$, and the hidden connectivity $\langle K_{\mbox{\scriptsize{h}}}\rangle$. The blue line corresponds to the critical curve when a single layer is observed while the red one refers to the whole system. The rest of simulation parameters are the same as for the other figures.}
\label{fig4}
\end{figure}

Up to now, we have analyzed the stability of the multiplex system. In practice, it is more common to have access to only one layer, so that one can measure the stability of that layer given that it is connected to a hidden (inaccessible) one. Therefore, it is also important to inspect the stability condition of a single layer within the multiplex. To this end, we should solve Eq. (\ref{lqc}) taking into account only the nodes that belong to the layer whose stability is scrutinized. In this case, the critical condition reads
\begin{equation}
\frac{\sigma}{4}(\langle K_{\mbox{\scriptsize{h}}}\rangle^2-\langle K_{\mbox{\scriptsize{o}}}\rangle^2+2\langle K_{\mbox{\scriptsize{h}}}\rangle\langle K_{\mbox{\scriptsize{o}}}\rangle)+\frac{\langle K_{\mbox{\scriptsize{o}}}\rangle^2}{2}=\frac{\langle K_{\mbox{\scriptsize{o}}}\rangle+\sigma\langle K_{\mbox{\scriptsize{h}}}\rangle}{2q}.
\label{single}
\end{equation}
Figure\ \ref{fig4} compares results of simulations for a larger network of $N=10^4$ nodes per layer with the theoretical solution (Eq.\ (\ref{single}), blue line) showing again a good agreement between analytical and simulation results. Remarkably, the results show that a single ingredient $-$the multilevel nature of the system $-$ can explain why there are biologically stable systems that are however theoretically expected to operate in the unstable regime (i.e., their average degree is larger than $1/q$). In other words, the sole reason could be that these systems are not isolated, but are coupled to other hidden layers that, if ordered, can stabilize the system. Finally, for the sake of comparisons, we have also represented in Fig.\ \ref{fig4} (red line) the case shown in Fig.\ \ref{fig2} but for the same larger system size.

Summing up, we have studied the effect of multiplexity on the stability of Boolean multilevel networks. In particular, we have addressed two important (and complementary) cases: the stability of the system as a whole and that of an observed layer which is coupled to other hidden layers. Our main result shows that there is a region of parameters for which either a single layer or the whole system can be stabilized by the presence of another stable sub-system (layer). On more general grounds, the latter mechanism supports the need to study complex interdependent systems explicitly incorporating their multilevel nature. As we have shown, unexpected results can emerge as a consequence of new ways of feedback between the structure and the dynamics of such systems, including the possibility of using interdependency to control the stability of a system.

\begin{acknowledgments}
This work has been partially supported by MINECO through Grants FIS2008-01240, FIS2009-13730-C02-02 and FIS2011-25167, by Comunidad de Arag\'on (Spain) through a grant to the group FENOL, and by Generalitat de Catalunya 2009-SGR-838. 
\end{acknowledgments}


\begin{thebibliography}{99}
\bibitem{note1} Throughout the manuscript we use the accent $(\tilde{})$ to denote quantities defined for the multiplex.
\bibitem{note3}  As a special case of interest, if $q_i=q$ and $\kappa_i=\kappa$, then $\lambda_Q=2^{-(\kappa-1)}q\lambda$ where $\lambda$ is the eigenvalue of the matrix sum of $A_{ij\alpha}$. Thus, the critical condition is $\lambda=\frac{2^{(\kappa-1)}}{q}$, and in the case of an uncorrelated network, the mean field approximation yields for the average degree of the multiplex $\langle  K \rangle =2^{\kappa-1}/q$.
\end{thebibliography}
\end{document}